\documentclass{vgtc}                          

\graphicspath{{figures/}{pictures/}{images/}{./}} 

\usepackage{times}                     

\usepackage{tabu}                      
\usepackage{booktabs}                  
\usepackage{lipsum}                    
\usepackage{mwe}                       

\usepackage{mathptmx}                  

\newcommand{\eg}{e.g.,~}

\onlineid{0}

\vgtccategory{Research}

\vgtcinsertpkg

\title{Gen4DS: Workshop on Data Storytelling in an Era of Generative AI}

\author{Xingyu Lan\thanks{e-mail: xingyulan96@gmail.com}\\ %
        \scriptsize Fudan University %
\and        Leni Yang\thanks{e-mail: yangleni96@gmail.com}\\ %
        \parbox{1.7in}{\scriptsize \centering Hong Kong University of Science and Technology} %
\and        Zezhong Wang\thanks{e-mail: wangzezhong2016@gmail.com}\\ %
        \scriptsize Simon Fraser University %
\and Yun Wang\thanks{e-mail: wangyun@microsoft.com}\\ %
     \scriptsize Microsoft Research Asia %
\\ \and Danqing Shi\thanks{e-mail: danqing.shi@aalto.fi}\\ %
         \scriptsize Aalto University %
  \and Sheelagh Carpendale\thanks{e-mail: sheelagh@sfu.ca}\\ %
         \scriptsize Simon Fraser University %
         }

\abstract{
    Storytelling is an ancient and precious human ability that has been rejuvenated in the digital age. Over the last decade, there has been a notable surge in the recognition and application of data storytelling, both in academia and industry.
    Recently, the rapid development of generative AI has brought new opportunities and challenges to this field, sparking numerous new questions. These questions may not necessarily be quickly transformed into papers, but we believe it is necessary to promptly discuss them to help the community better clarify important issues and research agendas for the future.
    We thus invite you to join our workshop (Gen4DS) to discuss questions such as: How can generative AI facilitate the creation of data stories? How might generative AI alter the workflow of data storytellers? What are the pitfalls and risks of incorporating AI in storytelling? We have designed both paper presentations and interactive activities (including hands-on creation, group discussion pods, and debates on controversial issues) for the workshop. We hope that participants will learn about the latest advances and pioneering work in data storytelling, engage in critical conversations with each other, and have an enjoyable, unforgettable, and meaningful experience at the event.
} 

\begin{document}


\firstsection{Motivation}

\maketitle

Over the last decade, the field of data storytelling has made significant progress. Researchers in the visualization community have conducted a series of important theoretical work, such as analyzing the narrative structure, visual design, and interactive design of data stories~\cite{segel2010narrative, hullman2013deeper,yang2021design}, constructing formative design spaces~\cite{brehmer2016timelines,bach2018design,shi2021communicating}, and evaluating the effects of data storytelling~\cite{wang2019comparing,lan2022negative}. 
At the same time, a series of tools to help create data stories have also been developed, including various interactive design editors (\eg Ellipsis~\cite{satyanarayan2014authoring} for authoring data articles, DataClips~\cite{amini2016authoring} for creating data videos) and intelligent tools to generate data stories automatically (\eg DataShot for generating fact sheets~\cite{wang2019datashot}, Calliope~\cite{shi2020calliope} for structured stories). What is perhaps more exciting is that the application of data storytelling has achieved significance both in academia and in the wild. For example, Segal and Heer's classic paper, \textit{Narrative Visualization: Telling Stories with Data}~\cite{segel2010narrative}, has been cited over two thousand times, with a considerable portion of the citations coming from other communities such as human-computer interaction, social science, business studies, and geography, serving the needs of industries such as news media, education, sports, politics, and the communication of natural science~\cite{tong2018storytelling,zhao2023stories}.

In the past, when creating data stories, whether writing narratives, creating graphics, or editing videos, a large amount of manual involvement was required. With the emergence and maturity of large models such as GPT, Gemini, Llama2, and Sora, AI-generated content is becoming the norm, and content production is undergoing drastic changes. This is full of opportunities and challenges for the field of data storytelling. 
On the one hand, with the emergence of many easy-to-use AI tools, the threshold and cost of creating data stories are significantly reduced. For example, now, ChatGPT can efficiently generate text descriptions and even complete stories, Midjourney can quickly generate illustrations, and tools like Pica and Runaway can quickly generate videos. Recently, Sora has made further breakthroughs in cross-modal generation between text, images, and videos, able to generate longer and clearer videos. More importantly, it demonstrates a strong ability to understand and narrate the world, such as changing perspectives, leveraging physical principles, and even expressing emotions~\cite{sora}. 
Undoubtedly, the aforementioned abilities will bring revolutionary changes to the creation and intelligent generation of data stories, having the potential to provide breakthroughs in some tough problems in previous research (\eg weak semantic understanding of data content in intelligent tools, lack of emotional expression in generated stories). 
On the other hand, the involvement of AI also brings new concerns. For example, the problem of blindly creating content by large models is a well-known risk~\cite{bender2021dangers}, which may cause serious problems in the intelligent generation of data stories. As stories are often presented to users in the form of insights or conclusions, it is difficult for users to return to the original data to reanalyze the conclusions if there is nonsense from AI. Furthermore, for the creators of data stories, what changes will there be in their workflows? Will there be resistance to AI? How should AI be integrated into their work and creation legally and ethically? These are all questions that need to be urgently discussed.

Hence, this workshop will center around the theme of "Data Storytelling in the Era of Generative AI (Gen4DS)," offering a platform to delve into the opportunities and challenges stemming from rapid technological advancements and illuminate the agenda for future research.
In general, we hope to adopt an open and dialogic approach to stimulate collective wisdom among community members. In addition to one-to-many presentations, we will also incorporate hands-on tutorials and group discussions to allow participants to experience and reflect on the impact of AI on data storytelling through hands-on activities and critical thinking. This approach aims to spark dynamics among participants, involving them not only as listeners but also as collaborators and contributors.

\section{Why Necessary}


The VIS community, with its dedication to visualization and visual analytics advancements, is well-placed to explore the integration of Generative AI in data-driven storytelling. The discussion of this integration can contribute to the current landscape and navigate the future trajectory of the field. The emergent technology, growing rapidly in 2024, is extending from professional to everyday applications, necessitating discourse within the VIS sphere on ethical, practical, and technological implications.

We believe that this workshop is significantly different from the main paper sessions at the VIS conference, and these two forms of activities can complement each other. The main reasons include:

\begin{itemize}
\setlength\itemsep{0em}
    \item As the technology of generative AI evolves extremely fast, a full paper, which requires significant effort and time to complete, may struggle to keep pace with the latest advancements and insights. By welcoming more ``light-weighted" papers, reports, and activities, the workshop can provide a timely platform for participants to share the most recent insights in generative AI and its implications for data storytelling.
    \item Secondly, compared to the formal paper programs, this workshop highly emphasizes discussion and interaction among participants. We will include a special hands-on creation session in the workshop. We have also innovatively designed a debate session focusing on AI controversies. We hope that these novel activities will enhance the enjoyment and engagement of participants during VIS and stimulate more meaningful insights.
    \item Thirdly, generative AI's cross-disciplinary nature highlights the need for an informal workshop setting at VIS, encouraging collaboration and idea exchange across fields like computer science, graphic design, and journalism. Such a setting promotes hands-on engagement with new tools, offering a platform for direct experimentation less feasible in formal conference settings.
    \item Last, this workshop is crucial for fostering a community among researchers, practitioners, and educators leading the charge in generative AI and data storytelling. It will facilitate knowledge sharing, debate, and multifaceted discussions, helping to navigate the ethical and innovative use of AI in storytelling, thereby making a significant contribution to the field's responsible advancement.
\end{itemize}

\section{Goals}

Firstly, we hope that participants in this workshop will have a fruitful experience, and this includes:

\setlength\itemsep{0em}
\noindent
\textbf{Exchanging.} Through paper presentations, participants will exchange their cutting-edge research and activities relating to data storytelling.

\noindent
\textbf{Exploring.} Through interactive activities, participants will explore how to integrate generative AI into the pipeline of data storytelling. The sessions are designed to foster an environment of collaborative learning, where participants can exchange insights on the latest tools, share their practical experiences, and discuss effective learning methodologies.

\noindent
\textbf{Criticizing.} By engaging in the discussion and debate on ``data storytelling + generative AI", participants are poised to uncover novel insights, deepen reflections, and draw inspiration. We also encourage participants to critically examine new technologies, guiding future research to utilize AI cautiously and responsibly.

\noindent
\textbf{Connecting.} Data storytelling is a highly interdisciplinary field, with researchers and practitioners coming from diverse backgrounds, organizations, and countries. By attending this workshop, participants will connect with more people with common interests and start meaningful conversations.

Secondly, we hope that this workshop can also yield long-term outcomes and spark continuous dialogues. This includes:

\setlength\itemsep{0em}

\noindent
\textbf{Synthesizing.} We plan to synthesize the valuable insights and opinions arising from the workshop and transform the collective wisdom into a white paper that suggests the research agenda and grand challenges of data storytelling in the era of strong artificial intelligence.

\noindent
\textbf{Researching.} We aim to conduct more in-depth research based on the interesting ideas and materials collected from the workshop and publish papers surrounding the topics in Section 3.

\noindent
\textbf{Building.} We will establish and maintain a long-term forum and interdisciplinary community centered around data storytelling, open to researchers, students, and practitioners.

\section{Scope of topics}

The workshop topics include but are not limited to:
\begin{itemize}
\setlength\itemsep{0em}
    \item Create multimedia content in data story with generative AI
    \item Data storytelling design workflows with generative AI
    \item Scenarios and application of data story with generative AI
    \item Human-AI collaboration mode in data story creation
    \item Pedagogy of data story creation with generative AI
    \item Ethics in generative AI for data storytelling
    \item Evaluation methods and criteria
    \item Usage guidelines of generative AI for data storytelling
    \item Tools/systems leveraging generative AI for data storytelling
    \item Data storytelling with generative AI for all
    \item Governance and autonomy 
\end{itemize}

\section{Submission Formats}
The workshop will accept four types of submissions, peer-reviewed by at least two PC members and one workshop organizer.

\begin{itemize}
\setlength\itemsep{0em}
    \item \textbf{Full papers:} 6-8 pages excluding references, intended for publication through the IEEE Xplore Digital Library with DOI. According to the guidance of IEEE VIS, accepted workshop papers can contribute to VIS full paper submissions in a future year. However, such submissions should not contain verbatim copies of previous content. Furthermore, any full paper submission to VIS must stand as its own contribution.
    \item \textbf{Short papers:} 2-4 pages excluding references, not for publication in the IEEE Xplore Digital Library; short papers will be available on the workshop’s website.
    \item \textbf{Activity reports:} practical reports that describe how an activity is conducted and how it could be reused by others and in other contexts; these activity reports will be published on the workshop's website.
    \item \textbf{Practical reports:} experiences and reflections on data storytelling experiences that serve the exchange between newly appointed and experienced faculty; these practical reports will be published on the workshop's website.
\end{itemize}

\section{Workshop activities}

We plan to have a half-day workshop with both paper presentations and more interactive activities. 

\subsection{Intended Format and Housekeeping Issues}

As this workshop includes hands-on creation and aims to encourage in-person deep interactions and discussions, it will be conducted in an offline format. Some housekeeping issues to consider for the offline format may include: 

\begin{itemize}
\setlength\itemsep{0em}
    \item \textbf{Venue Selection.} Securing an appropriate venue with adequate space and facilities for presentations and discussions.
    \item \textbf{Desk Arrangement.} Arranging the desks or seating into formats (\eg circles) that suit group discussions and collaborative activities, fostering an environment conducive to interaction and idea exchange.
    \item \textbf{Material Preparation.} Preparing and distributing workshop materials such as big-sized papers, colored pens, and sticky notes to help write down and organize ideas offered by group discussions.
    \item \textbf{On-site Support.} Designating organizers or volunteers to provide assistance and support to participants throughout the workshop.
\end{itemize}

\subsection{Activity Types}


\subsubsection{Paper Presentations} 
Each accepted paper will have approximately 8-10 minutes for presentation, with a maximum of 6-8 minutes for introducing the papers and the remaining time allocated for Q\&A.
The time left after all accepted papers have been presented will be flexibly allocated to the following interactive activities.

\subsubsection{Interactive Activities}
After the paper presentation, the interactive activities begin with experiencing through hands-on creation, then discussing and sharing experiences and opinions in group discussion pods that focus on different topics. Finally, a debate on topics of choice to engage in discussion and exchange opinions.

\textbf{Hands-on Creation:} 
We plan to engage the participants in experiencing some generative AI tools that have the potential to help the ideation and authoring of key elements in data storytelling. 
We will first have an onboarding session to introduce some mature and accessible generative AI tools. We will publish the learning materials ahead on the workshop website to help participants make necessary preparations.
Subsequently, participants will engage in data story creation to explore the integration of generative AI in the development of narrative structures and visualizations.
Participants will not be asked to create a whole data story but attempt to design some key elements of data stories, such as designing the narrative structures and visuals of critical scenes in data stories.
After that, we will collect participants' feedback on the challenges they encountered during accomplishing the tasks and how generative AI may affect their workflows in creating data stories. 
We expect this hands-on practice to inspire them for follow-up group discussions, and help participants leverage generative AI for their future work.

\textbf{Group Discussion Pods:} 
Based on the workshop topics and the accepted papers, we will provide a list of distinct aspects of generative AI in data storytelling for workshop participants to discuss future research opportunities and challenges regarding the combination of generative AI and data storytelling (e.g., ethics, collaboration modes, pedagogy).
We will have multiple small groups for the discussion. In the beginning, each group will look through the list of issues and can freely add new issues they identify as important.
Each discussion group will have a moderator who is expected to be one of the workshop organizers. The moderator will take charge of the time control, take notes of the discussion, and guide the discussion procedure. Using the approach of the Six Thinking Hats~\cite{carl1996six} encourages thinking from six distinct perspectives (i.e., Managing, Information, Emotions, Positive, and Creativity), to explore different aspects of a situation or problem thoroughly and collaboratively. Participants can choose to change their table during the discussion, they are encouraged to leave ideas and opinions on sticky notes on the table during discussion.
After the discussion ends, each group will summarize and report their conclusions in 2 minutes to all workshop participants.

\textbf{Debate Arena:} 
The last activity will feature a Debate Arena, where participants will engage in debates on controversial topics within the domain of generative AI and data storytelling. Participants can choose the topics they wish to debate and the side (proponent or opponent) they wish to represent. Grouping will be based on the number of participants choosing each topic and side, ensuring a balanced and engaging discussion. This segment is structured to include defined roles for proponents, opponents, and mediators, facilitating an informal debate process. The purpose of this arena is to foster critical discourse on complex issues, encouraging participants to articulate and examine diverse perspectives. This component of the workshop is designed to illuminate the multifaceted nature of ethical and practical considerations in the use of generative AI for data storytelling.


\subsection{Tentative Schedule}
The half-day workshop has the following tentative schedule.

\begin{itemize}
\setlength\itemsep{0em}
    \item 09:00 - 09:10 \textbf{Opening \& keynote}
    \item 09:10 - 09:50 \textbf{Paper presentation}
    \item 09:50 - 10:00 \textit{\textbf{Coffee break}}
    \item 10:00 - 10:30 \textbf{Hands-on creation}
    \item 10:30 - 11:00
    \textbf{Group discussion pods}
    \item 11:00 - 11:10 \textit{\textbf{Coffee break}}
    \item 11:10 - 11:20 \textbf{Discussion summary}
    \item 11:20 - 12:00 \textbf{Debate arena}
\end{itemize}

\subsection{Special Technology or Assistance Needed}

Although this workshop does not require special technology, for the hands-on tasks, we encourage each participant to bring a laptop and ensure smooth internet connectivity and browser performance. Additionally, since there will be on-site group discussions, to better track the viewpoints of each group and provide necessary support (\eg answering questions), we may need the assistance of 1-2 student volunteers.

\subsection{Intended Result and Impact}
The workshop will provide a channel for participants to discuss the opportunities and challenges brought by generative AI to data storytelling. Participants will present their latest research findings regarding leveraging generative AI for data storytelling, share their experiences and approaches to how generative AI can be better utilized, or exchange their perspectives on future research problems. 
We expect to receive more than 10 papers after the call for papers, with 3-6 papers to be presented at our workshop.

We will summarize the lessons learned about utilizing generative AI for data storytelling revealed in both paper presentations and discussion sessions. 
We will develop a website to publish our summary results and provide related materials for any researchers or data storytelling practitioners to use in future research and practices. 
We will conclude with a research agenda based on the discussion results. Based on this, we will continue to compile the latest research and cases from academia and industry to form a white paper. This will contribute to not only data storytelling researchers but also broader research communities such as AI-support creativity and human-AI collaboration.
We plan to submit new research related to the topics discussed in this workshop to visualization-related conferences or journals.

\section{Organization}

\subsection{Workshop Organization Timeline}
The timeline for the workshop organization is as follows:

\begin{itemize}
\setlength\itemsep{0em}
    \item March 20, 2024: Call for Participation
    \item July 1, 2024: Deadline for Paper Submission
    \item July 25, 2024: Reviews Collection
    \item July 30, 2024: Author Notification
    \item August 15, 2024: Camera Ready Deadline
\end{itemize}

\subsection{Development of the List of Participants}
For paper submissions, we plan to advertise on the respective mailing lists such as ACM
CHI, IEEE VIS, and social media (e.g., Twitter). For the discussion session, in addition to advertising, we will invite both researchers and practitioners actively involved in data storytelling, especially those focusing on the combination of AI and data storytelling.
The organizers, primarily young scholars, come from various countries and institutions and are enthusiastic and passionate about organizing the workshop.
Each of us has published research on data storytelling in top-tier venues such as IEEE TVCG and ACM CHI, covering topics such as narrative methods, visual and interactive design, evaluation, and the development of intelligent tools.
Our collaboration network will also aid in identifying potentially interested participants.

We value cultural diversity and believe that every voice matters. We will actively encourage individuals from all backgrounds to participate in the workshop and will strive to create a welcoming and respectful environment for all participants, regardless of their cultural or ethnic background.

\subsection{Backup Plan}
Our backup plan considers the situation of insufficient paper submissions or overrun of paper presentations.
For the former situation,
we will prepare enough sub-events in the three interactive activities to ensure that they can cover the whole workshop. 
For the hands-on activity, we will prepare at least three cases for participants to experience generative AI tools for data storytelling designs. 
At least one case will cover the basic usage of AI tools and others will involve more advanced usage. 
For the group discussion pods and debate arena activities, we will hold a brainstorming among workshop organizers to prepare topics. The topics in the papers submitted will be added into consideration further.
By manipulating the number of cases and topics for discussion and debate, we will have a flexibly changeable schedule to ensure participants can have a fruitful experience in this workshop.
On the other hand, if the schedule is too tight for interactive activities, we will prioritize the group discussion pods and consider shortening the hands-on creation and debate arena activities.


\section{Organizing Committee}

\textbf{Xingyu Lan}, xingyulan96@gmail.com

\noindent (\url{https://olivialan.github.io/})

\noindent Xingyu Lan is an assistant professor in the School of Journalism, Fudan University, Shanghai, China. Her research focuses on data storytelling, user experience, and human-AI interaction. She conducted a series of studies about users' affective responses to data stories~\cite{lan2021kineticharts,lan2021smile,lan2022negative,lan2021understanding} and received IEEE VIS Best Paper Award~\cite{lan2023affective}. 
She was once an award-winning practitioner and has rich experience in creating data stories.
She is now responsible for five courses concerning data visualization and HCI at Fudan University and has conducted several hands-on workshops with more than 300 participants for designing data stories using methods such as brainstorming, storyboarding, and design hackathon. She was also the co-organizer and host of several conference events, such as the Visualization Roast panel at the ChinaVis conference (2022, 2023), and the project manager of the Chinese visualization learning community, Gradict.\\

\noindent \textbf{Leni Yang}, yangleni96@gmail.com

\noindent (\url{https://yleni.github.io/})

\noindent Leni Yang is a post-doc research fellow in the Department of Computer Science and Engineering at the Hong Kong University of Science and Technology, Hong Kong SAR, China. Her research interests include Data Visualization and Human-Computer Interaction with a focus on investigating effective design theories, techniques, and authoring tools for turning complex data into understandable and compelling data stories~\cite{xu2022wow,yang2023understanding,yang2021design}. Her most recent publications are about immersive (XR) data storytelling and generative AI-assisted data story authoring. She has hosted several workshops where participants experience activities like designing storyboards for data stories, brainstorming, and group discussions. \\

\noindent \textbf{Zezhong Wang}, 
wangzezhong2016@gmail.com

\noindent (\url{https://zezhongwang.com/})

\noindent  Zezhong Wang is a postdoctoral researcher at the Interactive Experiences Lab, Simon Fraser University, Canada. His research interests encompass the creation and examination of data-driven storytelling, harnessing visual communication and insights derived from data, and employing design innovation to enhance data accessibility and engagement for the general public. Zezhong earned his PhD from the School of Informatics, University of Edinburgh, where he explored the creation of data comics. His research delved into the effectiveness of data comics~\cite{wang2019comparing}, their creation~\cite{wang2019teaching, wang2021reporting, wang2022interactive}, and materials for teaching visualization literacy~\cite{wang2020cheat}, earning him the VGTC Visualization Dissertation Award Honorable Mention in 2023. \\

\noindent \textbf{Yun Wang},
wangyun@microsoft.com

\noindent (\url{https://www.microsoft.com/en-us/research/people/wangyun/})

\noindent Yun Wang is a senior researcher in the Data, Knowledge, Intelligence (DKI) Area at Microsoft Research Asia. Her research interests lie at the intersection of human-computer interaction and information visualization, applying the advances in artificial intelligence and data science. Her work aims to facilitate human-data interaction, human-AI collaboration, visual storytelling, and natural language interfaces. She has developed techniques and systems for creating visual communication in diverse forms, such as infographics, interactive web pages, motion graphics, and animated videos. She has also explored how to simplify and improve the data interaction workflows between humans and AI for analysis, ideation, authoring, and storytelling (\eg ~\cite{wang2019datashot,wang2018infonice}). 
She has published more than 40 papers in high-impact venues such as VIS, CHI, UIST, TVCG, and CG\&A and serves as a reviewer and program committee for a variety of venues. \\

\noindent \textbf{Danqing Shi}, danqing.shi@aalto.fi

\noindent (\url{https://sdq.github.io/})

\noindent Danqing Shi is a postdoc researcher at the Finnish Center for Artificial Intelligence and Aalto University. His research interests include human-computer interaction and information visualization. He has strong experience in building automated visualization tools for data storytelling using AI-based approaches~\cite{shi2020calliope,shi2021autoclips}. Two related visualization systems for data storytelling developed by him have received IEEE VIS Best Poster Honorable Mention and ChinaVis Best Poster Honorable Mention~\cite{guo2021talk2data}. He has also co-hosted several workshops for data analysts on visualization design and visual data analysis. \\

\noindent \textbf{Sheelagh Carpendale}, sheelagh@sfu.ca

\noindent (\url{https://www.cs.sfu.ca/~sheelagh/})

\noindent Professor and Canada Research Chair in Data Visualization, at Simon Fraser University in the School of Computing Science is a Fellow of the Royal Society of Canada. Her many awards include the IEEE Visualization Career Award, CS-CAN Lifetime Achievement, a BAFTA and being inducted into both the ACM CHI (Computer-Human-Interaction) Academy and the IEEE Visualization Academy. As an international leader in information visualization and interaction design, she studies how people interact with information both in work and social settings, she works towards designing more natural, accessible and understandable interactive visual representations of data.

\bibliographystyle{abbrv-doi}

\bibliography{template}
\end{document}